\documentclass[final,5p,times,twocolumn]{elsarticle}

\usepackage{amssymb}
\usepackage{amsmath}
\usepackage{hyperref}
\usepackage{color}

\usepackage[latin]{armtex}

\journal{Physica C}

\begin{document}

\begin{frontmatter}



\title{Concealed Mott Criticality: \\ Unifying the Kondo Breakdown and Doped Charge-Transfer Insulators}


\author{Louk Rademaker} 

\affiliation{organization={Department of Quantum Matter Physics, University of Geneva},
            addressline={24 Quai Ernest-Ansermet}, 
            city={Geneva},
            postcode={1211}, 
            country={Switzerland}}
\affiliation{organization={Institute-Lorentz for Theoretical Physics, Leiden University},
            addressline={PO Box 9506}, 
            city={Leiden},
            postcode={NL-2300}, 
            country={The Netherlands}}

\begin{abstract}
I show that the quantum critical points observed in heavy fermions (the `Kondo breakdown') and in doped cuprates can be understood in terms of concealed Mott criticality. In this picture, one species of electrons undergoes a Mott localization transition, in the presence of metallic charge carries. As is shown in a simple toy model, this results in a Fermi surface jump at the transition, as well as mass enhancement on both the `large' and `small' Fermi surface side of the transition, consistent with the experimental observations.
\end{abstract}







\end{frontmatter}



\section{Introduction}
\label{Sec:Intro}

The field of strongly correlated electron systems historically developed from two seemingly different classes of materials: on the one hand the {\bf heavy fermion compounds}\cite{Löhneysen.2007,Rademaker.2016vc9,Paschen.2021,Hu.2024} such as CeAl$_3$, CeCu$_2$Si$_2$, and YbRh$_2$Si$_2$; and on the other hand the {\bf cuprates}\cite{Imada.1998,Zaanen.2006,Lee.200608b} such as La$_{2-x}$Sr$_x$CuO$_4$ and Bi$_2$Sr$_2$CaCu$_2$O$_{8+x}$.

Both classes exhibit 
unconventional `high-$T_c$' superconductivity, 
(antiferro)magnetic phases, 
linear-in-$T$ strange metallicity ($\rho \sim T$)\cite{Phillips.2022}, and signatures of quantum criticality\cite{Löhneysen.2007,Zaanen.2008}.
However, the theoretical starting point in attempts to describe them are radically different. 

Heavy fermions are considered to be captured by the {\bf Kondo lattice model}\cite{Coleman.2015p0w}
\begin{equation}
	H = \sum_{{\bf k} \sigma} \epsilon_{\bf k} c^\dagger_{{\bf k} \sigma} c_{{\bf k} \sigma}
		+ J_K \sum_i (c^\dagger_{i \alpha} \vec{\sigma}_{\alpha \beta} c_{i \beta}) \cdot \vec{S}_i
\end{equation}
that describes the coupling between conduction electrons $c_{{\bf k} \sigma}$ and a lattice of magnetic impurities $\vec{S}_i$. The behavior of the `heavy' Fermi liquid is viewed in terms of a many-impurity version of the Kondo effect\cite{Hewson.1993}. Tuning field or pressure causes a quantum phase transition between the heavy Fermi liquid and a magnetic phase. This transition, dubbed the `Kondo breakdown', is typically associated with a change in behavior of the spins of localized moments, following the seminal work of Doniach.\cite{Doniach.1977}

On the other hand, the starting point for cuprates is considered to be the single-band {\bf Hubbard model}\cite{Imada.1998}
\begin{equation}
	H = \sum_{{\bf k} \sigma} \epsilon_{\bf k} c^\dagger_{{\bf k} \sigma} c_{{\bf k} \sigma}
		+ U \sum_i n_{i \uparrow} n_{i \downarrow}
\end{equation}
which exhibits a Mott transition at half-filling\cite{Phillips.2006}. The physics of cuprates is therefore often viewed as that of `doping a Mott insulator': doping away from half-filling all the interesting phases appear such as the pseudogap, the strange metal and high temperature superconductivity.\cite{Lee.200608b,Phillips.2010} Recent developments in numerical techniques showed that the pseudogap phenomenology is caused by spin fluctuations.\cite{SimkovicIV.2024} However, underneath the superconducting dome there are signatures of a quantum critical point (QCP), whose origin is still hotly debated.

Despite the difference in understanding between cuprates and heavy fermions, it is important to notice that the QCP in both sets of materials has two major similarities. In both materials the transition is in between two metallic phases, characterized by:\cite{Hu.2024}
\begin{itemize}
\item A jump in the carrier density, as shown by Hall measurements or ARPES;\cite{Balakirev.2009,Badoux.2016,Fang.2022}
\item A divergence of the effective mass near the QCP.\cite{Ramshaw.2015lir}
\end{itemize}
In this paper, I show that both main features of the heavy fermion and doped cuprate QCP can be understood by a simple picture of having a Mott localization transition in the presence of additional metallic charge carriers. The {\bf Mott criticality} of the correlated electrons is then {\bf concealed} by the conduction electrons, yet the latter do inherit many properties of the Mott QCP including mass enhancement.



Indeed, we know that chemically the heavy fermions host multiple electron species, and rather than fractionalizing localized moments,\cite{Coleman.2015p0w} it is much more natural to interpret the `Kondo breakdown' as a localization transition, where on one side the $f$ electrons are itinerant and on the other side localized. Similarly, cuprates are actually not single-band materials but fall within a class of {\em charge-transfer insulators} following the Zaanen-Sawatzky-Allen classification.\cite{Zaanen.1985} A proper description involves both the correlated copper orbitals as well as the weakly correlated oxygen orbitals, known as the three-band Hubbard or Emery model.\cite{Emery.1987,Kung.2016} The hidden QCP could then be interpreted as a delocalization transition coming from the copper $d$ orbitals.

The challenge, however, is that Mott quantum criticality in itself is a theory in its infancy, despite the fact that the finite temperature Mott transition is known for more than half a century.\cite{Mott.1968,Mott.1990} Only recent theoretical works are exploring this,\cite{Misawa.2006,Terletska.2011,Vucicevic.2015,Eisenlohr.2019,Tan.2022,Takai.2023am7} though a quantitative agreement with experiments in organics and moiré systems is still lacking.\cite{Pustogow.2018,Li.202109b,Ghiotto.2021,Tan.2022} Most notably, within the numerical Dynamical Mean Field Theory (DMFT)\cite{Georges.1996}, the Mott transition is first-order, and therefore does not reproduce Mott criticality.

Instead of providing a detailed solution of the problem of Mott criticality, I approach this problem phenomenologically in Sec.~\ref{Sec:MottModel}. The resulting microscopic phenomenological theory of the Mott QCP is then applied to a toy model with two bands in Sec.~\ref{Sec:Model}, showing I can replicate the two main features of the heavy fermion/cuprate QCP: a jump in the Fermi surface and a diverging effective mass. I end with a brief discussion on the role of electron spin versus the role of the electron charge in Sec.~\ref{Sec:SpinDiscussion}.


Note that the idea that a Mott QCP can occur in systems with multiple orbitals is not new, it has been discussed in the context of ruthenates,\cite{Anisimov.2002} pnictides\cite{Medici.2008,Yu.2011,Yu.2012}, and moiré systems including twisted bilayer graphene\cite{Rademaker.20181uq,Rademaker.20198zm,Cea.2019qdv,Song.2022,Datta.2023,Rai.2024} and twisted transition-metal dichalcogenides\cite{Zhao.2023,Zhao.2024z7k,Brzezińska.2024,Xie.2024rmn}. A recent review paper on heavy fermions\cite{Hu.2024} even mentions in the outlook that cuprates and heavy fermions share a `hidden Mott transition'. It is my aim with this paper to make this connection more transparent.

\section{Phenomenology of Mott criticality}
\label{Sec:MottModel}

\begin{figure*}
	\includegraphics[width=\textwidth]{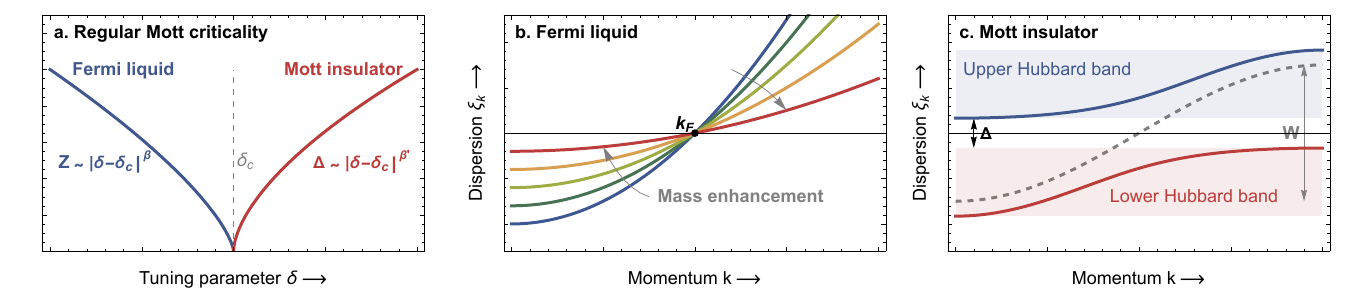}
	\caption{The simplified picture of regular Mott criticality. {\bf a.} Regular Mott criticality describes the continuous zero-temperature phase transition from a Fermi liquid to a Mott insulator in a single band system. On the Fermi liquid side, the quasiparticle weight $Z$ continuously goes to zero as a function of the tuning parameter $\delta$. On the Mott insulator side, the Mott gap $\Delta$ goes to zero as we approach the transition. {\bf b.} The effective quasiparticle dispersion is `flattened' due to the mass enhancement $m^*/m = 1/Z$ as we approach the Mott critical point from the Fermi liquid side. {\bf c.} On the Mott insulator side, the pole in the self-energy (Eq.~\eqref{Eq:MottSelfEnergy}) causes a split of the bare dispersion (gray dashed line) into an Upper and Lower Hubbard band.}
	\label{Fig:RegularMott}
\end{figure*}

The main perspective of this paper is that both the `Kondo breakdown' and the tentative quantum critical point in doped charge-transfer insulators can be understood in terms of a {\em hidden Mott critical point}. Our starting point should be, therefore, to discuss the properties of a `regular' single-band continuous Mott transition. At the moment there exists no microscopic theory of Mott criticality that is consistent with observations in 2DEGs, organics or moiré systems\cite{Tan.2022}. However, we can develop a {\em microscopic phenomenological theory} in the following sense: we will not attempt to derive the transition from ab initio or some given Hamiltonian; yet the theory will be microscopic in the sense that the core quantity we introduce is the (retarded) {\bf electron self-energy} $\Sigma({\bf k},\omega)$. This approach is very similar to, for example, the original work on the marginal Fermi liquid\cite{Varma.1989}, or recent work on the Kadowaki-Woods ratio\cite{Jacko.2009} or the optical conductivity scaling in cuprates\cite{Michon.2023}.

The question thus is: can we introduce an electron self-energy $\Sigma({\bf k},\omega)$ that accounts for a continuous Mott transition? For convenience, and inspired by the approach of DMFT\cite{Georges.1996}, we consider only a {\em local} self-energy $\Sigma(\omega)$, meaning independent of momentum. 



On the {\bf Fermi liquid} side of the Mott transition, the electron system should have a well-defined sharp Fermi surface. At zero temperature, an expansion in frequency close to the Fermi energy yields, to leading order,\cite{Imada.1998}
\begin{equation}
	\Sigma_{\rm FL} (\omega) = - (Z^{-1} - 1) \omega + \ldots
	\label{Eq:FLSelfEnergy}	
\end{equation}
where $Z$ is the quasiparticle weight, which is inversely proportional to the mass enhancement $\frac{m^*}{m} = Z^{-1}$. This can be seen from considering the poles of the Greens function $G({\bf k},\omega) = (\omega - \xi_{\bf k} - \Sigma_{\rm FL}(\omega))^{-1}$. Near the Fermi level, we can expand the non-interacting dispersion as $\xi_{\bf k} = \frac{k_F}{m} (k-k_F) + \ldots$. With the Fermi liquid self-energy of Eq.~\eqref{Eq:FLSelfEnergy}, the effective dispersion becomes
\begin{equation}
	\xi^*_k = Z \frac{k_F}{m} (k-k_F) + \ldots 
	\equiv \frac{k_F}{m^*} (k-k_F) + \ldots
\end{equation}
The Mott transition occurs when $Z$ continuously goes to zero, which correspond to a divergence of the effective mass. In general
\begin{equation}
	Z \sim | \delta - \delta_c|^\beta
\end{equation}
where $\delta$ is a tuning parameter (pressure, electric/magnetic field, doping, twist angle), $\delta_c$ is the value where the Mott quantum critical point occurs, and $\beta$ is a critical exponent that experimentally has been observed anywhere between 0.5 and 1.\cite{Tan.2022}



On the localized side of the transition, the {\bf Mott insulator} is characterized by having a pole in the self-energy,\cite{Phillips.2006}
\begin{equation}
	\Sigma_{\rm MI} (\omega) = \frac{\Delta W/4 }{\omega + i 0^+},
	\label{Eq:MottSelfEnergy}
\end{equation}
where $\Delta$ is the Mott gap and $W$ the non-interacting bandwidth. Indeed, one can verify that the diverging self-energy breaks the non-interacting bandwidth into an upper and lower Hubbard band. 
At each momentum, the Greens function $G({\bf k},\omega) = (\omega - \xi_{\bf k} - \Sigma_{\rm MI}(\omega))^{-1}$ is split into {\em two} poles, at respectively
\begin{equation}
	\xi^*_{\bf k,\pm} = \tfrac{1}{2} \xi_{\bf k} \pm \tfrac{1}{2} \sqrt{\xi_{\bf k}^2 + \Delta W}.
\end{equation}
The top of the lower Hubbard band is given by $\xi^*_{\bf k+}$ when $\xi_{\bf k} = +W/2$, which to leading order in $\Delta$ is $-\Delta/2$. Similarly, the bottom of the upper Hubbard band is located at $+\Delta/2$, given by $\xi^*_{\bf k-}$ when $\xi_{\bf k} = -W/2$. Therefore the total Mott gap is $\Delta$, provided $\Delta$ is smaller than the non-interacting bandwidth $W$. Of course, this condition is naturally satisfied near the Mott critical point when the gap continuously vanishes,
\begin{equation}
	\Delta \sim |\delta - \delta_c|^{\beta'}
\end{equation}
where $\beta'$ is a critical exponent, which in experiments has been seen to vary between 0.6 and 1.\cite{Tan.2022}

Note that the functional shape of the self-energy Eq.~\eqref{Eq:MottSelfEnergy} is the same as in DMFT\cite{Georges.1996}. However, in DMFT the strength of the self-energy pole remains nonzero as one approaches the transition, whereas for Mott criticality we require the pole strength to vanish continuously.

The phenomenology of `regular' single-band Mott quantum criticality is summarized in Fig.~\ref{Fig:RegularMott}. As a function of the tuning parameter $\delta$, either the quasiparticle weight or the Mott gap go to zero continuously. In terms of effective electron dispersions, this leads to either a mass enhancement or the formation of a upper and lower Hubbard band.

Note that I neglected many properties of the self-energy. In fact, above we only described the low-energy behavior of the real part of the local self-energy. The imaginary part is subleading, however, both on the Fermi liquid side (where it is responsible for scattering and the familiar $T^2$-resistivity) as well as on the Mott side (where it is responsible for making the upper/lower Hubbard bands incoherent). Similarly the momentum and possible spin-dependence of the self-energy is ignored. However, since we are interested in reproducing the relevant gap formation and mass enhancement, we do not need anything beyond Eqs.~\eqref{Eq:FLSelfEnergy}-\eqref{Eq:MottSelfEnergy}. It is time now to add a second, weakly interacting, electron species.

\section{Mott criticality with conduction electrons}
\label{Sec:Model}

Instead of a single band system with `regular' Mott criticality, we will now look at a system with {\em two} types of electrons. 
One type of electrons have a large bandwidth such that we can ignore their interactions, and we will refer to them as $c$ ("conduction") electrons.
The other type of electrons have a relative flat band compared to their interactions, and we will refer to them as $f$ ("flat") electrons. They will be endowed with the self-energy associated with Mott criticality as was discussed in the previous Section~\ref{Sec:MottModel}.

Additionally, we allow for a hybridization of the two electron types, which is identical to an inter-orbital hopping element between $c$ and $f$ electrons, parametrized by $t_\perp$. 

The interacting Greens function of the whole system will thus be a $2\times 2$ matrix in $c/f$ space,
\begin{equation}
	\hat{G}({\bf k}, \omega) = \begin{pmatrix}
		\omega - \xi_c ({\bf k}) + i 0^+ & - t_\perp \\ 
		-t_\perp & \omega - \xi_{f} ({\bf k}) - \Sigma_f (\omega) + i0^+
	\end{pmatrix}^{-1}.
	\label{Eq:GreensF}
\end{equation}
I will now show that this model can account for the two main experimentally observed characteristics of the `Kondo breakdown' and the QCP in cuprates, namely:
\begin{itemize}
\item A jump in the size of the Fermi surface, or equivalently a jump in the number of charge carriers;
\item A diverging mass enhancement (a `heavy Fermi liquid') as we approach the transition.
\end{itemize}

In order to make it slightly more concrete and to allow for visualizations, I consider the following non-interacting bandstructure. For the correlated $f$ electrons, we take a nearest-neighbor square lattice dispersion,
\begin{equation}
	\xi_{f} ({\bf k}) = -t (\cos k_x + \cos k_y)
\end{equation}
where $t >0$ is the hopping, such that the bandwidth is $W=4t$. With $\mu=0$ the $f$ electrons are fixed at half-filling. The conduction $c$ electrons should have a total density that is less than half-filling, so we ignore the lattice and consider a parabolic dispersion with a light mass,
\begin{equation}
	\xi_c ({\bf k}) = \frac{k^2}{2m} - \mu_c.
\end{equation}
The corresponding density of $c$ electrons is $n_c = \frac{m \mu}{\pi}$. 

In the absence of any interactions on the $f$ electrons, the whole system is described by two bands that are of mixed $c/f$-orbital character, with dispersion
\begin{equation}
	\xi_{h, \pm} ({\bf k}) =  \frac{1}{2} \left( \xi_c({\bf k}) + \xi_f ({\bf k}) 
	\pm \sqrt{ \left( \xi_c({\bf k}) - \xi_f ({\bf k}) \right)^2  + 4 t_\perp^2} \right)
	\label{Eq:HybridizedDispersion}
\end{equation}
In the visualizations that follow I choose the parameters $t=1$, $t_\perp = 3$, $m = 0.05$ and $\mu =3$. I chose $t_\perp > \sqrt{2 t \mu}$ so that the hybridized band-structure has only one large Fermi surface. The main results with regard to the Fermi surface jump and the mass enhancement are not dependent on the precise parameters chosen.

\begin{figure}
	\includegraphics[width=\columnwidth]{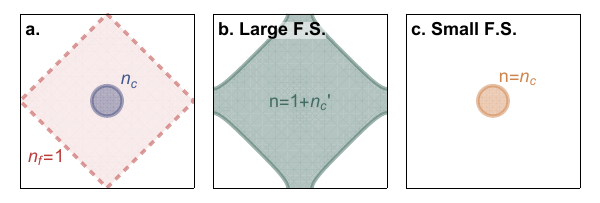}
	\caption{The Fermi surfaces in the different phases of our model. {\bf a.} In the absence of interactions and the inter-orbital hopping $t_\perp$, there is a half-filled Fermi surface associated with the $f$ electrons and a small Fermi surface for the $c$ electrons. {\bf b.} If the $f$ electrons are in the Fermi liquid regime, the resulting bandstructure has a single large Fermi surface with a volume given by electron density $n = 1+n_c'$. {\bf c.} When the $f$ electrons localize, only the small Fermi surface remains with density given by the $c$-electron density. }
	\label{Fig:FermiSurfaces}
\end{figure}

\subsection{Jump in the Fermi surface}

In any metallic system, the Fermi surface is defined as the momenta ${\bf k}$ where the Greens function has a pole at zero frequency, $\omega=0$. 

On the Fermi liquid side of our interacting model, the self-energy $\Sigma_f(\omega)$ vanishes at $\omega = 0$, and the system's Fermi surface is defined through the non-interacting hybridized dispersion, Eq.~\eqref{Eq:HybridizedDispersion}, by setting $\xi_{h, -} ({\bf k}) = 0$. The Fermi surface size corresponds to a electron density larger than 1, which I will denote by $n = 1+n_c'$. Note that $n_c'$ is in this simplified model not equal to $n_c$, as the hybridization parameter $t_\perp$ causes an effective shift in chemical potential for the hybridized bands. For now, we ignore this shift, and focus on the difference between the Fermi liquid and Mott insulator.

On the Mott insulator side, when the $f$ electrons are localized, something interesting is happening. As we approach $\omega = 0$, the self-energy $\Sigma_f(\omega)$ diverges. As a result, the $c$ and $f$ electrons effectively decouple, and we find that the Fermi surface is the one of only the $c$ electrons.
Note that this decoupling only happens at the Fermi level $\omega = 0$, and should in no way be confused with a change in the hybridization parameter $t_\perp$.

Combining these results we find that at the transition, there is a jump of a Fermi surface with size $1+n_c'$ to $n_c$. This is indeed what has been observed in heavy fermions at the Kondo breakdown, cuprates near the QCP, and moiré systems. In this simplified model, the `concealed' Mott criticality of the $f$ electrons automatically causes this jump. This picture is summarized in Fig.~\ref{Fig:FermiSurfaces}.

\subsection{Mass enhancement on `large' side}

\begin{figure}
	\includegraphics[width=\columnwidth]{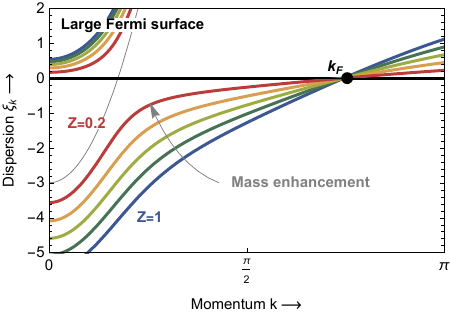}
	\caption{In the regime with a large Fermi surface, the hybridized band structure given by Eq.~\eqref{Eq:LargeFSBS} inherits the mass enhancement of the $f$ electrons. This figure shows the dispersions for $Z =0.2$ to $Z=1$ in steps of $0.2$.}
	\label{Fig:FLDispersion}
\end{figure}

To study the mass enhancement, we need to take one step further and extract the poles of the full Greens function of Eq.~\eqref{Eq:GreensF} at any frequency $\omega$. These poles are found by solving the following equation,
\begin{equation}
	(\omega - \xi_c({\bf k}) ( \omega - \xi_f ({\bf k}) - \Sigma_f (\omega)) - t_\perp^2 = 0.
	\label{Eq:PolesGreensF}
\end{equation}
In the case of the Fermi liquid regime, the effect of the $f$ self-energy is to introduce mass enhancement via $\omega - \Sigma_f(\omega) = Z^{-1} \omega$. The hybridized bands in the `large Fermi surface' regime are thus
\begin{equation}
	\xi_{\rm large} ({\bf k}) = \frac{1}{2} \left( 
	\xi_c({\bf k}) + Z \xi_f ({\bf k}) 
	- \sqrt{ \left( \xi_c({\bf k}) -Z  \xi_f ({\bf k}) \right)^2 + 4 Zt_\perp^2}\right)
	\label{Eq:LargeFSBS}
\end{equation}
To find the mass enhancement in this regime, we need to linearize the dispersion around the Fermi momentum $k_F$. In general, as can be seen in Fig.~\ref{Fig:FermiSurfaces}b, the Fermi momentum depends on the direction $\theta$. The Fermi momentum $k_F(\theta)$ will in general be where the bare $c$ and $f$ dispersions $\xi_c({\bf k})$ and $\xi_f({\bf k})$ are nonzero. We expand these bare dispersions, in the direction perpendicular to the large Fermi surface, as $\xi_c ({\bf k}) = \xi_c^{(0)} + \xi_c^{(1)} (k - k_F^{({\rm big})})$ and $\xi_f ({\bf k}) = \xi_f^{(0)} + \xi_f^{(1)} (k - k_F^{({\rm big})})$. In terms of this expansion, the hybridized dispersion becomes
\begin{equation}
	\xi_{\rm big} ({\bf k} \approx k_F) = Z \frac{\xi_c^{(1)} \xi_f^{(0)} + \xi_c^{(0)} \xi_f^{(1)}}{\xi_f^{(0)} + Z \xi_f^{(0)}} (k-k_F) + \ldots
\end{equation}
Interestingly, the effective mass {\em diverges} in the same way as for regular Mott criticality! The hybridized large Fermi surface {\em inherits} the mass enhancement from the $f$-electrons, with the same critical exponent,
\begin{equation}
	m^*_{\rm large} \sim Z^{-1} \sim |\delta - \delta_c|^{-\beta}.
\end{equation}
This mass enhancement is visualized in Fig.~\ref{Fig:FLDispersion}.

\subsection{Mass enhancement on `small' side}

\begin{figure}
	\includegraphics[width=\columnwidth]{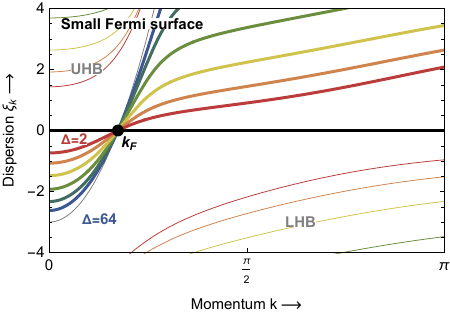}
	\caption{When there is a small Fermi surface due to the Mott localization of the $f$ electrons, the small Fermi surface exhibits a mass enhancement proportional to the inverse of the Mott gap $\Delta$. This figure shows the dispersions for a range of gap values of the form $\Delta = 2^n$, from $n=2$ to 64. In the limit of an infinite Mott gap, the $c$ electron dispersion is regained. The thin lines indicate the lower and upper Hubbard bands associated with the $f$ electrons.}
	\label{Fig:MottDispersion}
\end{figure}

To get the dispersion in the regime with a small Fermi surface, we need to solve the pole equation~\eqref{Eq:PolesGreensF} with the Mott self-energy for the $f$ electrons. This constitutes a cubic polynomial, which has exact solutions that are cumbersome to write down. Since we are only interested in the mass enhancement, we instead linearize the pole equation near $k_F$, where $k_F$ is the Fermi momentum of the $c$ electrons. This yields
\begin{equation}
	\xi_{\rm small} ({\bf k} \approx k_F) = \frac{\Delta}{t_\perp^2/t + \Delta} v_F (k-k_F) + \cdots
\end{equation}
where $k_F, v_F$ are the Fermi momentum and Fermi velocity of the $c$ electrons {\em without} hybridization. Again, for a small $f$ Mott gap $\Delta$ near the transition, we find an effective mass that diverges, with the divergence exponent inherited from hidden Mott criticality,
\begin{equation}
	m^*_{\rm small} \sim \frac{t_\perp^2}{t\Delta} \sim | \delta - \delta_c|^{- \beta'}.
\end{equation}
This mass enhancement is visualized in Fig.~\ref{Fig:MottDispersion}.

\subsection{Inherited criticality}

We have seen that assuming a `concealed' Mott criticality of the $f$ electrons results in a `large' and `small' Fermi surface regime, with a Fermi surface jump at the transition. The effective mass in these metallic states diverges as we approach the transition from either side. This shows that the critical behavior of this metal-to-metal transition is inherited from the underlying Mott metal-to-insulator transition.

Similarly, but beyond the scope of this work, is that {\em at} the transition there are scaling forms for the self-energy and susceptibilities, which again will be inherited from the concealed Mott criticality of the $f$ electrons.

\section{The role of spin}
\label{Sec:SpinDiscussion}

The perspective I present here neglects many things which could affect the critical behavior, most importantly the spin of the electron. Of course, without the spin {\em degeneracy} of the electron, a Mott transition cannot be possible, not even a `concealed' one. The analysis of the previous section, however, shows that the spin {\em degrees of freedom} play a secondary role, in that their dynamics are not required to cause a Fermi surface jump nor mass enhancement. 

This view is at odds with the prevailing `Doniach' dogma\cite{Doniach.1977}, in which RKKY interactions and antiferromagnetic order tune the Kondo breakdown transition. It is similarly contrary to the idea that the cuprate QCP is related to the collapse of the pseudogap, which recently has been unequivocally connected to spin fluctuations.\cite{SimkovicIV.2024}

The approach of Sec.~\ref{Sec:Model} can be summarized as follows: {\em charge first, spin second}. This is a subtle point, because the moment the local moments develop, it is likely some form of magnetic order or fluctuations follow. And in general the opposite scenario (spin first, charge second) can also exist: a spin-density wave instability can gap parts of the Fermi surface. However, in that scenario, the underlying transition is akin to a simple band transition\cite{Fratini.2023}, which does {\em not} exhibit any mass enhancement. 

When the Kondo breakdown is treated on a mean-field level, fractionalization of a local spin moment is required.\cite{Coleman.2015p0w} The charge `appears' as the result of fractionalizing spin, which seems unnecessary, when one realizes that in the end we are dealing with (charged) electrons anyway. The effective fractionalized Hamiltonian is (often) the same as an original microscopic theory, if we interpret the Schwinger fermions `$f$' as just the original $f$-electrons. There is one subtle caveat, though: in most parton mean field theories of the Kondo breakdown, the effective order parameter that tunes the transition is the emergent hybridization (the `Kondo singlet formation').\cite{Coleman.2015p0w} However, in the perspective of this paper, the hybridization between $f$ and $c$ electrons remains fixed, and it is purely the localization of the physical $f$ electrons that causes the metal-to-metal transition.

\section{Outlook}
\label{Sec:Outlook}

In conclusion, I have shown that the main features of the QCP in the cuprates and heavy fermions -- the Fermi surface jump and the mass enhancement -- can be interpreted in terms of `concealed' Mott criticality in the presence of charge carriers.

In addition to neglecting spin (discussed in Sec.~\ref{Sec:SpinDiscussion}), there are many other aspects that might affect the critical behavior. This includes, but is not limited to, the precise number of orbitals, their degeneracies and their degree of correlations, the different possible couplings between the orbitals, the manner of tuning through the transition, and the topology of the underlying bandstructure. Much of these different aspects can be tuned within moiré systems,\cite{Rademaker.20181uq,Rademaker.20198zm,Cea.2019qdv,Song.2022,Datta.2023,Rai.2024,Zhao.2023,Zhao.2024z7k,Brzezińska.2024,Xie.2024rmn}. My hope is by studying moiré systems in quantitative detail, we can learn more about correlated metal-to-metal QCPs.

Finally, a question that comes to mind is: so what? What do we gain from this `concealed Mott' interpretation? In my opinion, it means that the big correlated questions -- the origin of strange metallicity and high temperature superconductivity -- should be studied in terms of a concealed Mott transition. The challenge is that there is currently no theoretical description of `regular' Mott criticality that is consistent with experimental observations.\cite{Tan.2022} Therefore, we need to go beyond the existing approaches (including dynamical and parton mean field theories) to capture `regular' Mott criticality, so that we can understand `concealed' Mott criticality and the exotic behavior of cuprates and heavy fermions.

\appendix
\section*{Acknowledgements}

I acknowledge discussions with Ajit Srivastava, Qimiao Si, Christophe Berthod, and Simone Fratini.
In particular, I want to mention the various discussions with my former PhD supervisor Jan Zaanen. He always imposed on me the importance of looking beyond mean field theories and non-interacting band structures, as well as the important of being critical of commonly held wisdoms, and never to forget experimental and chemical reality.
This work was funded by the Swiss National Science Foundation (SNSF) via Starting Grant TMSGI2\_211296.


\begin{thebibliography}{10}
\expandafter\ifx\csname url\endcsname\relax
  \def\url#1{\texttt{#1}}\fi
\expandafter\ifx\csname urlprefix\endcsname\relax\def\urlprefix{URL }\fi
\expandafter\ifx\csname href\endcsname\relax
  \def\href#1#2{#2} \def\path#1{#1}\fi

\bibitem{Löhneysen.2007}
H.~v. Löhneysen, A.~Rosch, M.~Vojta, P.~Wölfle,
  \href{http://link.aps.org/doi/10.1103/RevModPhys.79.1015}{{Fermi-liquid
  instabilities at magnetic quantum phase transitions}}, Reviews of Modern
  Physics 79~(3) (2007) 1015 -- 1075.
\newblock \href {https://doi.org/10.1103/revmodphys.79.1015}
  {\path{doi:10.1103/revmodphys.79.1015}}.

\bibitem{Rademaker.2016vc9}
L.~Rademaker, J.~Mydosh, {Chapter 280 Quantum Critical Matter and Phase
  Transitions in Rare Earths and Actinides}, Handbook on the Physics and
  Chemistry of Rare Earths 49 (2016) 293--338.
\newblock \href {https://doi.org/10.1016/bs.hpcre.2016.03.002}
  {\path{doi:10.1016/bs.hpcre.2016.03.002}}.

\bibitem{Paschen.2021}
S.~Paschen, Q.~Si, \href{http://dx.doi.org/10.1038/s42254-020-00262-6}{{Quantum
  phases driven by strong correlations}}, Nature Reviews Physics 3~(1) (2021)
  9--26.
\newblock \href {https://doi.org/10.1038/s42254-020-00262-6}
  {\path{doi:10.1038/s42254-020-00262-6}}.

\bibitem{Hu.2024}
H.~Hu, L.~Chen, Q.~Si, {Quantum critical metals and loss of quasiparticles},
  Nature Physics 20~(12) (2024) 1863--1873.
\newblock \href {https://doi.org/10.1038/s41567-024-02679-7}
  {\path{doi:10.1038/s41567-024-02679-7}}.

\bibitem{Imada.1998}
M.~Imada, A.~Fujimori, Y.~Tokura,
  \href{https://link.aps.org/doi/10.1103/RevModPhys.70.1039}{{Metal-insulator
  transitions}}, Reviews of Modern Physics 70~(4) (1998) 1039 -- 1263.
\newblock \href {https://doi.org/10.1103/revmodphys.70.1039}
  {\path{doi:10.1103/revmodphys.70.1039}}.

\bibitem{Zaanen.2006}
J.~Zaanen, S.~Chakravarty, T.~Senthil, P.~W. Anderson, P.~Lee, J.~Schmalian,
  M.~Imada, D.~Pines, M.~Randeria, C.~Varma, M.~Vojta, M.~Rice,
  \href{http://www.nature.com/articles/nphys253}{{Towards a complete theory of
  high Tc}}, Nature Physics 2~(3) (2006) 138 -- 143.
\newblock \href {https://doi.org/10.1038/nphys253}
  {\path{doi:10.1038/nphys253}}.

\bibitem{Lee.200608b}
P.~A. Lee, N.~Nagaosa, X.-G. Wen,
  \href{https://link.aps.org/doi/10.1103/RevModPhys.78.17}{{Doping a Mott
  insulator: Physics of high-temperature superconductivity}}, Reviews of Modern
  Physics 78~(1) (2006) 17 -- 85.
\newblock \href {https://doi.org/10.1103/revmodphys.78.17}
  {\path{doi:10.1103/revmodphys.78.17}}.

\bibitem{Phillips.2022}
P.~W. Phillips, N.~E. Hussey, P.~Abbamonte, {Stranger than metals}, Science
  377~(6602) (2022) eabh4273.
\newblock \href {https://doi.org/10.1126/science.abh4273}
  {\path{doi:10.1126/science.abh4273}}.

\bibitem{Zaanen.2008}
J.~Zaanen,
  \href{http://www.sciencemag.org/cgi/doi/10.1126/science.1152443}{{Quantum
  Critical Electron Systems: The Uncharted Sign Worlds}}, Science 319~(5867)
  (2008) 1205 -- 1207.
\newblock \href {https://doi.org/10.1126/science.1152443}
  {\path{doi:10.1126/science.1152443}}.

\bibitem{Coleman.2015p0w}
P.~Coleman, \href{https://books.google.ch/books?id=kcrZCgAAQBAJ}{{Introduction
  to Many-Body Physics}}, Cambridge University Press, 2015.

\bibitem{Hewson.1993}
A.~C. Hewson, {The Kondo Problem to Heavy Fermions}, Cambridge University
  Press, Cambridge University Press, 1993.

\bibitem{Doniach.1977}
S.~Doniach,
  \href{http://adsabs.harvard.edu/cgi-bin/nph-data\_query?bibcode=1977PhyBC..91..231D\&link\_type=EJOURNAL}{{The
  Kondo lattice and weak antiferromagnetism}}, Physica B \& C 91 (1977) 231 --
  234.
\newblock \href {https://doi.org/10.1016/0378-4363(77)90190-5}
  {\path{doi:10.1016/0378-4363(77)90190-5}}.

\bibitem{Phillips.2006}
P.~Phillips,
  \href{https://linkinghub.elsevier.com/retrieve/pii/S0003491606000765}{{Mottness}},
  Annalen der Physik 321~(7) (2006) 1634 -- 1650.
\newblock \href {https://doi.org/10.1016/j.aop.2006.04.003}
  {\path{doi:10.1016/j.aop.2006.04.003}}.

\bibitem{Phillips.2010}
P.~Phillips,
  \href{https://link.aps.org/doi/10.1103/RevModPhys.82.1719}{{Colloquium:
  Identifying the propagating charge modes in doped Mott insulators}}, Reviews
  of Modern Physics 82~(2) (2010) 1719 -- 1742.
\newblock \href {https://doi.org/10.1103/revmodphys.82.1719}
  {\path{doi:10.1103/revmodphys.82.1719}}.

\bibitem{SimkovicIV.2024}
F.~ŠimkovicIV, R.~Rossi, A.~Georges, M.~Ferrero, {Origin and fate of the
  pseudogap in the doped Hubbard model}, Science 385~(6715) (2024) eade9194.
\newblock \href {https://doi.org/10.1126/science.ade9194}
  {\path{doi:10.1126/science.ade9194}}.

\bibitem{Balakirev.2009}
F.~F. Balakirev, J.~B. Betts, A.~Migliori, I.~Tsukada, Y.~Ando, G.~S.
  Boebinger, {Quantum Phase Transition in the Magnetic-Field-Induced Normal
  State of Optimum-Doped High-Tc Cuprate Superconductors at Low Temperatures},
  Physical Review Letters 102~(1) (2009) 017004.
\newblock \href {https://doi.org/10.1103/physrevlett.102.017004}
  {\path{doi:10.1103/physrevlett.102.017004}}.

\bibitem{Badoux.2016}
S.~Badoux, W.~Tabis, F.~Laliberté, G.~Grissonnanche, B.~Vignolle,
  D.~Vignolles, J.~Béard, D.~A. Bonn, W.~N. Hardy, R.~Liang,
  N.~Doiron-Leyraud, L.~Taillefer, C.~Proust, {Change of carrier density at the
  pseudogap critical point of a cuprate superconductor}, Nature 531~(7593)
  (2016) 210--214.
\newblock \href {https://doi.org/10.1038/nature16983}
  {\path{doi:10.1038/nature16983}}.

\bibitem{Fang.2022}
Y.~Fang, G.~Grissonnanche, A.~Legros, S.~Verret, F.~Laliberté, C.~Collignon,
  A.~Ataei, M.~Dion, J.~Zhou, D.~Graf, M.~J. Lawler, P.~A. Goddard,
  L.~Taillefer, B.~J. Ramshaw, {Fermi surface transformation at the pseudogap
  critical point of a cuprate superconductor}, Nature Physics 18~(5) (2022)
  558--564.
\newblock \href {https://doi.org/10.1038/s41567-022-01514-1}
  {\path{doi:10.1038/s41567-022-01514-1}}.

\bibitem{Ramshaw.2015lir}
B.~J. Ramshaw, S.~E. Sebastian, R.~D. McDonald, J.~Day, B.~S. Tan, Z.~Zhu,
  J.~B. Betts, R.~Liang, D.~A. Bonn, W.~N. Hardy, N.~Harrison, {Quasiparticle
  mass enhancement approaching optimal doping in a high-Tc superconductor},
  Science 348~(6232) (2015) 317--320.
\newblock \href {https://doi.org/10.1126/science.aaa4990}
  {\path{doi:10.1126/science.aaa4990}}.

\bibitem{Zaanen.1985}
J.~Zaanen, G.~A. Sawatzky, J.~W. Allen,
  \href{https://link.aps.org/doi/10.1103/PhysRevLett.55.418}{{Band gaps and
  electronic structure of transition-metal compounds}}, Physical Review Letters
  55~(4) (1985) 418 -- 421.
\newblock \href {https://doi.org/10.1103/physrevlett.55.418}
  {\path{doi:10.1103/physrevlett.55.418}}.

\bibitem{Emery.1987}
V.~J. Emery, {Theory of high-Tc superconductivity in oxides}, Physical Review
  Letters 58~(26) (1987) 2794--2797.
\newblock \href {https://doi.org/10.1103/physrevlett.58.2794}
  {\path{doi:10.1103/physrevlett.58.2794}}.

\bibitem{Kung.2016}
Y.~F. Kung, C.~C. Chen, Y.~Wang, E.~W. Huang, E.~A. Nowadnick, B.~Moritz, R.~T.
  Scalettar, S.~Johnston, T.~P. Devereaux,
  \href{https://link.aps.org/doi/10.1103/PhysRevB.93.155166}{{Characterizing
  the three-orbital Hubbard model with determinant quantum Monte Carlo}},
  Physical Review B 93~(15) (2016) 155166 -- 14.
\newblock \href {https://doi.org/10.1103/physrevb.93.155166}
  {\path{doi:10.1103/physrevb.93.155166}}.

\bibitem{Mott.1968}
N.~F. Mott, {Metal-Insulator Transition}, Reviews of Modern Physics 40~(4)
  (1968) 677--683.
\newblock \href {https://doi.org/10.1103/revmodphys.40.677}
  {\path{doi:10.1103/revmodphys.40.677}}.

\bibitem{Mott.1990}
N.~Mott, \href{https://books.google.ca/books?id=Q0mJQgAACAAJ}{{Metal-Insulator
  Transitions}}, Taylor \& Francis, Taylor \& Francis, 1990.

\bibitem{Misawa.2006}
T.~Misawa, M.~Imada, {Quantum criticality around metal-insulator transitions of
  strongly correlated electron systems}, Physical Review B 75~(11) (2006)
  115121.
\newblock \href {https://doi.org/10.1103/physrevb.75.115121}
  {\path{doi:10.1103/physrevb.75.115121}}.

\bibitem{Terletska.2011}
H.~Terletska, J.~Vucicevic, D.~Tanasković, V.~Dobrosavljević,
  \href{http://adsabs.harvard.edu/cgi-bin/nph-data\_query?bibcode=2011PhRvL.107b6401T\&link\_type=EJOURNAL}{{Quantum
  Critical Transport near the Mott Transition}}, Physical Review Letters
  107~(2) (2011) 026401.
\newblock \href {https://doi.org/10.1103/physrevlett.107.026401}
  {\path{doi:10.1103/physrevlett.107.026401}}.

\bibitem{Vucicevic.2015}
J.~Vucicevic, D.~Tanasković, M.~J. Rozenberg, V.~Dobrosavljević,
  \href{https://link.aps.org/doi/10.1103/PhysRevLett.114.246402}{{Bad-Metal
  Behavior Reveals Mott Quantum Criticality in Doped Hubbard Models}}, Physical
  Review Letters 114~(24) (2015) 246402 -- 5.
\newblock \href {https://doi.org/10.1103/physrevlett.114.246402}
  {\path{doi:10.1103/physrevlett.114.246402}}.

\bibitem{Eisenlohr.2019}
H.~Eisenlohr, S.-S.~B. Lee, M.~Vojta,
  \href{http://arxiv.org/abs/1906.05293v1}{{Mott quantum criticality in the
  one-band Hubbard model: Dynamical mean-field theory, power-law spectra, and
  scaling}}, Physical Review B 100~(15) (2019) 155152.
\newblock \href {https://doi.org/10.1103/physrevb.100.155152}
  {\path{doi:10.1103/physrevb.100.155152}}.

\bibitem{Tan.2022}
Y.~Tan, V.~Dobrosavljevic, L.~Rademaker, {How to Recognize the Universal
  Aspects of Mott Criticality?}, Crystals 12 (2022) 932.
\newblock \href {https://doi.org/10.3390/cryst12070932}
  {\path{doi:10.3390/cryst12070932}}.

\bibitem{Takai.2023am7}
K.~Takai, Y.~Yamaji, F.~F. Assaad, M.~Imada, {Quantum criticality of
  bandwidth-controlled Mott transition}, Physical Review Research 5~(3) (2023)
  033186.
\newblock \href {https://doi.org/10.1103/physrevresearch.5.033186}
  {\path{doi:10.1103/physrevresearch.5.033186}}.

\bibitem{Pustogow.2018}
A.~Pustogow, M.~Bories, A.~Löhle, R.~Rösslhuber, E.~Zhukova, B.~Gorshunov,
  S.~Tomić, J.~A. Schlueter, R.~Hübner, T.~Hiramatsu, Y.~Yoshida, G.~Saito,
  R.~Kato, T.~H. Lee, V.~Dobrosavljević, S.~Fratini, M.~Dressel,
  \href{http://dx.doi.org/10.1038/s41563-018-0140-3}{{Quantum spin liquids
  unveil the genuine Mott state}}, Nature Materials 17~(9) (2018) 1 -- 6.
\newblock \href {https://doi.org/10.1038/s41563-018-0140-3}
  {\path{doi:10.1038/s41563-018-0140-3}}.

\bibitem{Li.202109b}
T.~Li, S.~Jiang, L.~Li, Y.~Zhang, K.~Kang, J.~Zhu, K.~Watanabe, T.~Taniguchi,
  D.~Chowdhury, L.~Fu, J.~Shan, K.~F. Mak,
  \href{http://dx.doi.org/10.1038/s41586-021-03853-0}{{Continuous Mott
  transition in semiconductor moiré superlattices}}, Nature 597 (2021) 350.
\newblock \href {https://doi.org/10.1038/s41586-021-03853-0}
  {\path{doi:10.1038/s41586-021-03853-0}}.

\bibitem{Ghiotto.2021}
A.~Ghiotto, E.-M. Shih, G.~S. S.~G. Pereira, D.~A. Rhodes, B.~Kim, J.~Zang,
  A.~J. Millis, K.~Watanabe, T.~Taniguchi, J.~C. Hone, L.~Wang, C.~R. Dean,
  A.~N. Pasupathy,
  \href{https://www.nature.com/articles/s41586-021-03815-6}{{Quantum
  Criticality in Twisted Transition Metal Dichalcogenides}}, Nature 597 (2021)
  345.
\newblock \href {https://doi.org/10.1038/s41586-021-03815-6}
  {\path{doi:10.1038/s41586-021-03815-6}}.

\bibitem{Georges.1996}
A.~Georges, G.~Kotliar, W.~Krauth, M.~J. Rozenberg,
  \href{http://link.aps.org/doi/10.1103/RevModPhys.68.13}{{Dynamical mean-field
  theory of strongly correlated fermion systems and the limit of infinite
  dimensions}}, Reviews of Modern Physics 68~(1) (1996) 13 -- 125.
\newblock \href {https://doi.org/10.1103/revmodphys.68.13}
  {\path{doi:10.1103/revmodphys.68.13}}.

\bibitem{Anisimov.2002}
V.~Anisimov, I.~Nekrasov, D.~Kondakov, T.~Rice, M.~Sigrist, {Orbital-selective
  Mott-insulator transition in Ca2 - xSrxRuO4}, The European Physical Journal B
  - Condensed Matter and Complex Systems 25~(2) (2002) 191--201.
\newblock \href {https://doi.org/10.1140/epjb/e20020021}
  {\path{doi:10.1140/epjb/e20020021}}.

\bibitem{Medici.2008}
L.~d. Medici, S.~R. Hassan, M.~Capone, X.~Dai, {Orbital-Selective Mott
  Transition out of Band Degeneracy Lifting}, Physical Review Letters 102~(12)
  (2008) 126401.
\newblock \href {https://doi.org/10.1103/physrevlett.102.126401}
  {\path{doi:10.1103/physrevlett.102.126401}}.

\bibitem{Yu.2011}
R.~Yu, Q.~Si, {Mott transition in multiorbital models for iron pnictides},
  Physical Review B 84~(23) (2011) 235115.
\newblock \href {https://doi.org/10.1103/physrevb.84.235115}
  {\path{doi:10.1103/physrevb.84.235115}}.

\bibitem{Yu.2012}
R.~Yu, Q.~Si, {U(1) slave-spin theory and its application to Mott transition in
  a multiorbital model for iron pnictides}, Physical Review B 86~(8) (2012)
  085104.
\newblock \href {https://doi.org/10.1103/physrevb.86.085104}
  {\path{doi:10.1103/physrevb.86.085104}}.

\bibitem{Rademaker.20181uq}
L.~Rademaker, P.~Mellado, \href{arXiv.org}{{Charge-transfer insulation in
  twisted bilayer graphene}}, Physical Review B 98~(23) (2018) 235158.
\newblock \href {https://doi.org/10.1103/physrevb.98.235158}
  {\path{doi:10.1103/physrevb.98.235158}}.

\bibitem{Rademaker.20198zm}
L.~Rademaker, D.~A. Abanin, P.~Mellado,
  \href{https://link.aps.org/doi/10.1103/PhysRevB.100.205114}{{Charge
  smoothening and band flattening due to Hartree corrections in twisted bilayer
  graphene}}, Physical Review B 100~(20) (2019) 205114.
\newblock \href {https://doi.org/10.1103/physrevb.100.205114}
  {\path{doi:10.1103/physrevb.100.205114}}.

\bibitem{Cea.2019qdv}
T.~Cea, N.~R. Walet, F.~Guinea,
  \href{https://journals.aps.org/prb/abstract/10.1103/PhysRevB.100.205113}{{Electronic
  band structure and pinning of Fermi energy to Van Hove singularities in
  twisted bilayer graphene: A self-consistent approach}}, Physical Review B 100
  (2019) 205113.
\newblock \href {https://doi.org/10.1103/physrevb.100.205113}
  {\path{doi:10.1103/physrevb.100.205113}}.

\bibitem{Song.2022}
Z.-D. Song, B.~A. Bernevig, {Magic-Angle Twisted Bilayer Graphene as a
  Topological Heavy Fermion Problem}, Physical Review Letters 129~(4) (2022)
  047601.
\newblock \href {https://doi.org/10.1103/physrevlett.129.047601}
  {\path{doi:10.1103/physrevlett.129.047601}}.

\bibitem{Datta.2023}
A.~Datta, M.~J. Calderón, A.~Camjayi, E.~Bascones, {Heavy quasiparticles and
  cascades without symmetry breaking in twisted bilayer graphene}, Nature
  Communications 14~(1) (2023) 5036.
\newblock \href {https://doi.org/10.1038/s41467-023-40754-4}
  {\path{doi:10.1038/s41467-023-40754-4}}.

\bibitem{Rai.2024}
G.~Rai, L.~Crippa, D.~Călugăru, H.~Hu, F.~Paoletti, L.~d. Medici, A.~Georges,
  B.~A. Bernevig, R.~Valentí, G.~Sangiovanni, T.~Wehling, {Dynamical
  Correlations and Order in Magic-Angle Twisted Bilayer Graphene}, Physical
  Review X 14~(3) (2024) 031045.
\newblock \href {https://doi.org/10.1103/physrevx.14.031045}
  {\path{doi:10.1103/physrevx.14.031045}}.

\bibitem{Zhao.2023}
W.~Zhao, B.~Shen, Z.~Tao, Z.~Han, K.~Kang, K.~Watanabe, T.~Taniguchi, K.~F.
  Mak, J.~Shan, {Gate-tunable heavy fermions in a moiré Kondo lattice}, Nature
  616~(7955) (2023) 61--65.
\newblock \href {https://doi.org/10.1038/s41586-023-05800-7}
  {\path{doi:10.1038/s41586-023-05800-7}}.

\bibitem{Zhao.2024z7k}
W.~Zhao, B.~Shen, Z.~Tao, S.~Kim, P.~Knüppel, Z.~Han, Y.~Zhang, K.~Watanabe,
  T.~Taniguchi, D.~Chowdhury, J.~Shan, K.~F. Mak, {Emergence of ferromagnetism
  at the onset of moiré Kondo breakdown}, Nature Physics 20~(11) (2024)
  1772--1777.
\newblock \href {https://doi.org/10.1038/s41567-024-02636-4}
  {\path{doi:10.1038/s41567-024-02636-4}}.

\bibitem{Brzezińska.2024}
M.~Brzezińska, S.~Grytsiuk, M.~Rösner, M.~Gibertini, L.~Rademaker,
  {Pressure-tuned many-body phases through $\Gamma$-K valleytronics in moiré
  bilayer WSe\$\_2\$}, arXiv (2024).
\newblock \href {https://doi.org/10.48550/arxiv.2404.07165}
  {\path{doi:10.48550/arxiv.2404.07165}}.

\bibitem{Xie.2024rmn}
F.~Xie, L.~Chen, Q.~Si, {Kondo effect and its destruction in heterobilayer
  transition metal dichalcogenides}, Physical Review Research 6~(1) (2024)
  013219.
\newblock \href {https://doi.org/10.1103/physrevresearch.6.013219}
  {\path{doi:10.1103/physrevresearch.6.013219}}.

\bibitem{Varma.1989}
C.~M. Varma, P.~B. Littlewood, S.~Schmitt-Rink, E.~Abrahams, A.~E. Ruckenstein,
  {Phenomenology of the normal state of Cu-O high-temperature superconductors},
  Physical Review Letters 63~(18) (1989) 1996--1999.
\newblock \href {https://doi.org/10.1103/physrevlett.63.1996}
  {\path{doi:10.1103/physrevlett.63.1996}}.

\bibitem{Jacko.2009}
A.~C. Jacko, J.~O. Fjærestad, B.~J. Powell, {A unified explanation of the
  Kadowaki–Woods ratio in strongly correlated metals}, Nature Physics 5~(6)
  (2009) 422--425.
\newblock \href {https://doi.org/10.1038/nphys1249}
  {\path{doi:10.1038/nphys1249}}.

\bibitem{Michon.2023}
B.~Michon, C.~Berthod, C.~W. Rischau, A.~Ataei, L.~Chen, S.~Komiya, S.~Ono,
  L.~Taillefer, D.~v.~d. Marel, A.~Georges, {Reconciling scaling of the optical
  conductivity of cuprate superconductors with Planckian resistivity and
  specific heat}, Nature Communications 14~(1) (2023) 3033.
\newblock \href {https://doi.org/10.1038/s41467-023-38762-5}
  {\path{doi:10.1038/s41467-023-38762-5}}.

\bibitem{Fratini.2023}
S.~Fratini, S.~Ciuchi, V.~Dobrosavljević, L.~Rademaker, {Universal Scaling
  near Band-Tuned Metal-Insulator Phase Transitions}, Physical Review Letters
  131~(19) (2023) 196303.
\newblock \href {https://doi.org/10.1103/physrevlett.131.196303}
  {\path{doi:10.1103/physrevlett.131.196303}}.

\end{thebibliography}



\end{document}